# Beyond a dichotomic approach to concepts: the case of colour phenomena[1]

L. Viennot[(1)] & C. de Hosson[(1)]

[(1)]Laboratoire de didactique André Revuz,
PRES Sorbonne Paris Cité
Université Paris Diderot-Paris 7

**Abstract:** *The research documents the impact of a teaching experiment designed in order to make students consider not only the spectral composition of light but also its intensity when they interpret colour phenomena. The analysis of eight recorded and transcribed teaching interviews conducted with third-year university students indicates that the comprehension of the colour vision process benefits from a "more or less" approach. Such an approach is likely to facilitate students' understanding of everyday observed phenomena.*

Background, framework and purpose

It is commonly advocated that, in order to teach concepts, we need to provide students with a simplified access to the real world. This process may pose little danger for an appropriate learning of physics if the descriptive models and statements chosen remain consistent with the observable facts. But this is not always the case and some 'teaching rituals' (Viennot, 2006) are sometimes based on prototypical statements contrary to what can daily be observed by students. "Black objects do not reflect any light" is one of these inadequate statements that can be found in some teaching materials within the learning of light sources context[2]. This statement contradicts the fact that the impact of a red laser beam remains visible whatever the colour of the surface on which the light is sent could be. Indeed, asserting that "black objects do not reflect any light" leans on an oversimplified view of the real world involving an 'all or nothing' reasoning process and does not include the fact that light should be considered in a quantitative way. Such an oversimplified way of considering light phenomena can be observed in teaching rituals concerning colour vision. The classical rules of additive mixing provide for instance the outcome of two beams of coloured lights superimposed on a white screen (e.g. "red plus green result yellow") correspondingly, the outcome of sending a coloured light beam on a filter or a pigment may be given by a rule of subtractive mixing such as "red plus yellow gives red". But such a reduced stating of the rules does not permit a proper interpretation of what students can observe in daily life. This is not only for practical reasons (for instance an omnipresent ambient light) but also because of the need to consider both the composition of the light sent and its intensity. In this research, we present the impact of a teaching pathway designed to stress the process of colour vision, pathway that takes into account the intensity of light. Following, in particular, the work of de Hosson and Kaminski (2007), our intension is to lead students from an 'all or nothing' way of reasoning to another one in 'more or less' terms, in order to make the visibility of the impact of a laser beam on any coloured surface intelligible. Our main research question is focused on the concept of absorption in the context of colour phenomena, concept that has proved to be very difficult (Chauvet, 2006).

---

[1] Texte de la communication présenté à la conférence ESERA 2011.
[2] http://maths-sciences.fr/documents/quatrieme/sources-de-lumiere-4eme.pdf, http://www.sciences92.ac-versailles.fr/spip/spip.php?article27. Links verified on December 27$^{th}$ of 2010.

Rationale

We seek to know to which extent students with the usual background concerning colour can benefit from a teaching pathway designed to stress the role of both light intensity and spectral composition. Our investigation is set in the framework of the *didactical engineering* (Artigue, 1994) where some hypotheses on the expected teaching learning processes can be tested through the confrontation of an *a priori* and an *a posteriori* analysis. In order to refine the *a priori* analysis, we chose to conduct a preliminary investigation using the *teaching experiment* method (Komorek & Duit 2004), which implies a discussion orientated towards conceptual acquisition, one that is strongly structured and guided, and allows students to evidence their initial thoughts and their reactions to diverse questions and requests.

Methodology of the research

The impact of the didactic intervention was measured through the analysis of 8 recorded interviews conducted with 8 prospective physics and chemistry teachers (third-year University) considered as very likely having been taught colour phenomena in a ritual way (basic rules of additive and subtractive colour mixing).The guided pathway can be described as follows: the student is first reminded of the rule of additive mixing of lights (e.g. red + green → yellow) in a context where its works, that is in a black room and with light beams of similar intensities. A setting of coloured shadows with three lamps (Olivieri *et al.*, 1988, Chauvet, 1996) is shown to demonstrate, with a simple equipment, all of the basic rules of additive and subtractive mixing, and the student is invited to predict the outcome of various changes in this setting. These changes are only on the register of presence/absence (phase P/A) of a given coloured beam. Doing so, he/she is provided with the table of the classical rules, in order to avoid any problem of memory. The second phase of the interviews is the critical one (phase M/L: more or less). The light source is now a laser pointer, and what is to be predicted, then observed, is the effect of this intense beam on papers printed with the six basic pigments of the graphics palette of a computer (red, green, blue, yellow, cyan and magenta) with a very weak ambient light. The most productive phase of the discussion is expected to occur when, for instance, the impact on a green paper is red (more details are given in the next section). The intellectual path we intend to facilitate should lead the student to realize that some predictions are nevertheless possible, provided the concept of partial absorption is used.

The interviews transcripts were submitted to a thematic content analysis, each category being defined by a theme that can be identified in the students' comments. The data is processed using two types of categories. Some originate in our a priori analysis and others emerge from the transcripts (Strauss & Corbin 1990). We pinpoint and count the occurrence of each category. This process was conducted independently by the two authors and the final classification emerged from a negotiation between them.

Results

The phase of prediction in the context of change by presence/absence (P/A) is just enacted to remind the interviewees with old memories, introduce the table of rules, and put the students in a situation of using successfully these tables. No particular difficulty was encountered. The second phase (M/L) was itself divided in two parts the results of which are as follows. In the first part of this phase (M/L1), it was observed that, concerning the pigments commonly described as "absorbing red light" (black, blue, green, and cyan), a majority of students (6/8) predicted that the impact of the laser beam would not be visible on the background. The first allusion to the intensity of the beam (3/8) happened in the case of a red paper illuminated by the (red) laser beam, that is, interestingly, in a case when the common rules predicted no absorption. In such a case, students (5/8) seemed to be freer to envisage "more or less" diffused (red) light. In the cases with magenta and yellow pigments, students provided complex reasoning involving the idea of mixing of colour, be it in a naïve way (like with paints: 2/8; see Chauvet 1996) or expert way (involving the common rules and the ambient light: 1/8). It was also observed that a given student might well keep to the same type of reason throughout the interview, e.g. "it's my experience (that the impact will be red)". In the second part of this critical phase (M/L2), students were invited to put their prediction to the test. Beyond the often unpredicted colour of the impact –red – what emerged clearly for all of the students was the brightness observed : different for two groups of pigments. A first group is constituted of the pigments that commonly are said to absorb red lights: black, blue, green and cyan. The second group of cases corresponds to the "non red-absorbing" pigments (red, magenta and yellow). In order to discuss this evidence of "more or less absorption", students were presented with the passing bands of various filters, and invited to comment on the "feet" of the corresponding curves. In this abstract, we only mention two important results: at the end of this teaching interview all of the students were able to express their conceptual progression, by contrasting a presence/absence perspective with a "more or less" approach. Moreover they all expressed a notable intellectual satisfaction (Viennot, 2006) linked - we suggest - to the fact that finally the experimental facts, although surprising, could be explained on a rational basis, in particular thanks to the curves they were presented with.

## Conclusion and implications

As always with teaching interviews, it is difficult to control all the components of conceptual input on behalf of the interviewer. Here, at least one of these inputs revealed to be essential: the discussion initiated about the absorption curves and their "feet". It is not very risky to assert that, without such documents and accompanying comments, the surprise generated by the observations described above would not have constituted a decisive factor. Only the soft edges of the passing bands and their multiplicative meaning, could help reconcile the common rules (P/A), the observed impacts and a "more or less" approach of absorption. In that sense, if such teaching interviews where called on in support of an Inquiry Based Science Education program, it would be essential not to forget this essential teacher intervention. It is our intention to conduct other investigations about this "more or less" approach, in contrast with a presence/absence view. This is not only because it is closer to everyday experience, but also because it may provide a facilitated access to the concept itself.